# Parallel magnetic field induced magnetoresistance peculiarities of the double quantum well filled with electrons or holes

M.V. Yakunin[1*], G.A. Alshanskii[1], Yu.G. Arapov[1], G.I. Harus[1], V.N. Neverov[1],
N.G. Shelushinina[1], O.A. Kuznetsov[2], B.N. Zvonkov[2], E.A. Uskova[2],
L. Ponomarenko[3], A. deVisser[3]

[1] *Institute of Metal Physics RAS, Ekaterinburg 620219, Russia*
[2] *Physico-Technical Institute at Nizhnii Novgorod State University, Nizhnii Novgorod, Russia*
[3] *Van der Waals - Zeeman Institute, University of Amsterdam, The Netherlands*

**Abstract**

In $In_xGa_{1-x}As/n$-GaAs double quantum wells (DQWs) containing an *electron* gas, the magnetoresistance (MR) peculiarities under parallel magnetic fields caused by the passing of the tunnel gap edges through the Fermi level are revealed. Peculiarities positioned in high fields (~30 T) can only be explained if the spin-splitting of the $In_xGa_{1-x}As$ conduction band is considered, that was neglected in the GaAs/AlGaAs heterostructures, for which solely the effects of this nature have been observed so far. In the $Ge/p$-$Ge_{1-x}Si_x$ DQWs containing a *hole* gas, local MR peculiarities under parallel fields are discovered as well. But the tunnel gap in these DQWs is too narrow to be responsible for these observations. We suppose, they are due to a complicated shape of the *hole* confinement subbands.

## 1. Introduction

A magnetic field configured parallel to the quasi-2D layer, $B = B_x$, causes (i) a diamagnetic shift of the energy levels, so that the distances between them increase, and (ii) a shift along $k_y$ of the energy dispersion surfaces $E_i(k_\parallel)$, $k_\parallel = (k_x, k_y)$ [1]. The latter is essential for a system of two coupled layers since their $E_i^{1,2}(k_\parallel)$ surfaces shift relative each other on $\Delta k_y^{1,2} = eBd/\hbar$, $d$ – an effective interlayer distance [2]. A tunnel gap $\Delta_{SAS}$ existing in the energy spectrum of this double quantum well (DQW) is fixed to the intersection line of the two $E_i(k_\parallel)$ paraboloids. The upper edge of $\Delta_{SAS}$ corresponds to a minimum $E_m$ of the inner surface in the joint energy dispersion, formed by the laterally shifted paraboloids, and the lower edge – to a saddle point $E_s$ on the external surface (fig.2). The gap rises in energy as the paraboloids laterally move away from each other with increasing field.

If the zero-field Fermi level position $E_F$ is above the gap, the $E_m$ minimum would cross the Fermi level at a field $B_m$, and then a saddle point cross it at a field $B_s > B_m$. At $B > B_m$ the density of states (DOS) at the Fermi level drops down and the intersubband scattering is quenched resulting in a down step or a minimum in magnetoresistivity (MR) $\rho(B_\parallel)$ at $B \approx B_m$. Also, a van Hove divergence in DOS is connected with the saddle point, causing a MR maximum at $B \approx B_s$. Peculiarities of both types have been found in DQW created in the conduction band of the traditional GaAs/n-AlGaAs heterosystem [3-5].

---

[*]Corresponding author. E-mail: yakunin@imp.uran.ru

In this paper we report on investigations of the DQWs under parallel magnetic fields in materials other than GaAs/n-AlGaAs. First, we present results for $In_xGa_{1-x}As/n$-GaAs DQW containing the *electron* gas. At least one new property is important here – a significant spin splitting of the conduction band due to a *g*-factor value of the InAs component, which is ~35 times larger than in GaAs. Second are the results for $Ge/p$-$Ge_{1-x}Si_x$ DQW containing a *hole* gas. A novel situation occurs in the valence band DQW due to a mixing of the heavy and light hole states.

## 2. $In_xGa_{1-x}As/n$-GaAs DQW

$In_{0.18}Ga_{0.82}As/n$-GaAs DQWs consist of 5 nm wide wells and GaAs barriers symmetrically silicon doped with 19 nm spacers on both sides of DQW. Two samples have been studied: 2981(2984) with the barrier width of 7(3.5) nm, electron gas density $n_s = 2.05(2.34)\cdot 10^{15}$ m$^{-2}$ and low-temperature mobility of 2.6(1.6) m$^2$/V·s. A rich quantum Hall (QH) structure present for both samples in perpendicular fields. Results in parallel fields are depicted in fig.1. The potential profile, energy levels and wave functions are determined from self-consistent calculations of the Schrodinger and Poison equations for zero magnetic field: $\Delta_{SAS} = 7.4(23.1)$ meV and the Fermi level $E_F = 8(9.5)$ meV. Since $E_F > \Delta_{SAS}$ in sample 2981, both a minimum and a maximum exist in its MR at fields $B_m$ and $B_s$, respectively. Contrary, $E_F \ll \Delta_{SAS}$ in sample 2984, and only a maximum exists at $B_s \approx 30$ T. (fig.1).

The DQW energy dispersion under parallel magnetic field is found from the two-level problem:

$$E_{1,2} = \frac{\hbar^2(k_x^2 + k_y^2)}{2m} + \frac{H_{ss} + H_{aa}}{2} \pm \frac{1}{2}\sqrt{(H_{ss} - H_{aa})^2 + 4H_{sa}^2}, \quad (1)$$

with $H_{ss} = E_s + \frac{m}{2}\omega^2 \langle s | z^2 | s \rangle$,

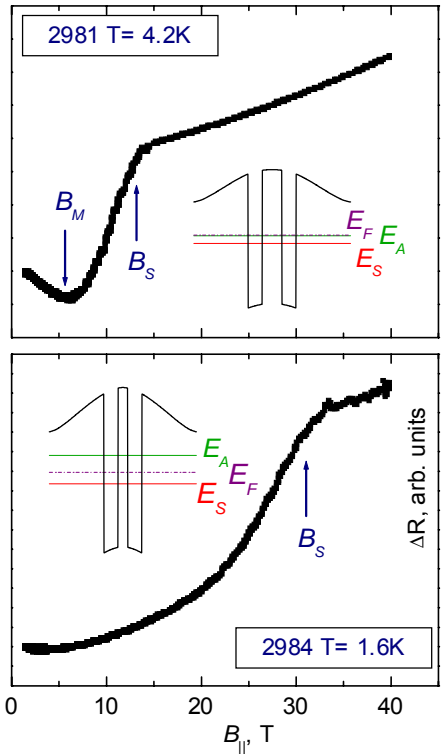

Fig.1. MR of the InGaAs/n-GaAs DQWs. Inserts: calculated potential profiles and energy levels.

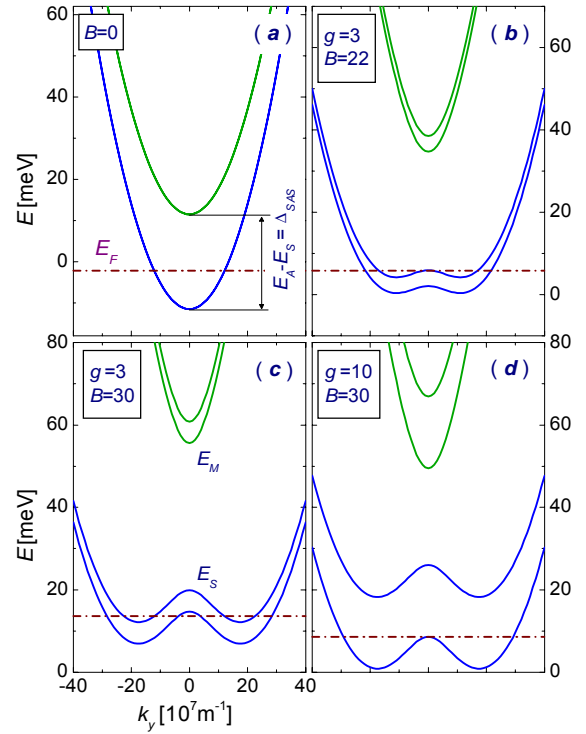

Fig.2. Energy dispersion along $k_y$ of the narrow barrier sample 2984 at different fields and *g*-factor values (indicated).

$$H_{aa} = E_a + \frac{m}{2}\omega^2 \langle a | z^2 | a \rangle,$$

$H_{sa} = -\hbar\omega k_y \langle s | z | a \rangle$, $\omega = eB/m$, $m$ – electron effective mass, $|s\rangle$ and $|a\rangle$ - symmetric and asymmetric DQW wavefunctions, $E_s$ and $E_a$ – lower and upper gap edges in zero field.

The simplest approach was used in [2] with $\langle s | z^2 | s \rangle = \langle a | z^2 | a \rangle = 0$. In this case the gap $H_{aa} - H_{ss} = E_a - E_s = \Delta_{SAS}$ is magnetic field independent. This approach is rather accurate to describe the $B_s$ position in our wide barrier sample 2981 (fig.3, top) but it is insufficient for the case of a large gap in the thin barrier sample 2984 (fig.3, bottom). Exact calculations indicate that in fact the gap increase with field reaching about twice the zero field value at $B = 30$ T (fig.2). Results of exact calculation satisfy the position of MR maximum for sample 2984 provided the spin splitting of the $In_{0.18}Ga_{0.82}As$ conduction band is considered with the $g$-factor $|g| > \sim 3$ [for this case the $\pm g\mu_B B/2$ terms should be added to (1)]. Interpolation between InAs and GaAs ($g = -15$ and $g = -0.44$, respectively) yields $g = -3$ for $In_{0.18}Ga_{0.82}As$.

Comparison of figures 2b and 2c yields the possible reason why the upper spin-subband doesn't show up in the experiment. At $B_m = 22$ T the saddle point hasn't resolved yet from the lateral minima on the same curve and the DOS peak is smeared, while at $B_s = 30$ T they do resolve. Moreover, in the latter case at $|g| \approx 3$ the saddle point of the lower spin subband coincides in energy with the lateral minima of the upper spin subband that lead to an enhancement of the resulting DOS peak and of corresponding MR maximum.

Comparison of figures 2c and 2d yields the explanation why the $B_s$ position saturates with increasing $|g|$ (fig.3): after a depopulation of the upper spin-subband the Fermi level becomes rigidly fixed within the lower one.

## 3. Ge/*p*-GeSi self formed DQW

We have found some local peculiarities in the parallel magnetic field MR for the Ge/*p*-GeSi heterostructures containing the *hole* gas in a self-formed DQW within a Ge layer (samples 475/476 in fig.4). The DQW potential profile emerges in a Ge layer wider than $d_w \approx 30$ nm at hole densities above $p_s \approx 1 \cdot 10^{15}$ m$^{-2}$ (fig.5), that manifests in the disappearance of the QH plateaus for filling factor $\nu = 1$ [6]. The MR traces for samples of the same heterosystem but with more narrow Ge layers (~20 nm, samples 1123 and 1125 in fig.4), for which the DQW profile has not formed yet, are smooth. Thus it is tempting to explain the peculiarities in the wide layer samples in analogy with those observed in the DQW with electron gas. But our calculations indicate that the tunnel gap here is very narrow, $\Delta_{SAS} \sim 0.1$ meV, due to a large hole mass. The magnetic field interval $B_s - B_m$ for such a narrow $\Delta_{SAS}$ should be about 0.1 T, more than order of magnitude smaller than those observed experimentally (fig.4, insert).

In a parallel magnetic field the samples with a self-formed DQW reveal a strong negative MR similar to that observed in samples 1123 and 1125. The obvious cause for the negative MR in the latter samples is a depopulation of the second subband due to its diamagnetic shift. In samples 475/476 the two lower subbands merge in the DQW energy profile, but the *third* subband is populated (fig.5) if the hole mass in the lower subbands is essentially different in directions perpendicular and parallel to the layer: $m_\perp/m_\parallel > \sim 5$. This condition is feasible for the valence band due to a mixing of the heavy and light hole states [7], and low values for $m_\parallel = (0.1\text{-}0.14)\, m_0$, while the bulk mass $m_\perp = 0.5\, m_0$, are indeed determined from the temperature damping of the Shubnikov – de Haas oscillations.

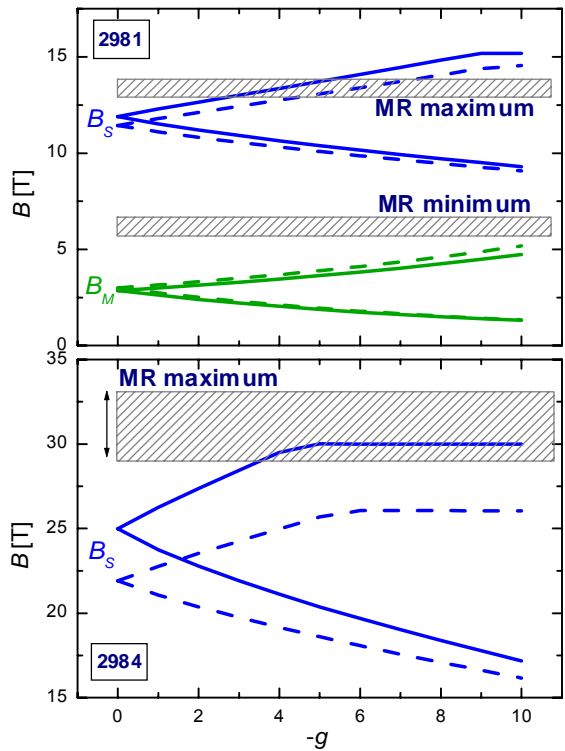

Fig.3. Comparison of the experimental MR maxima and minima positions (hatched areas) with calculations. Dashed lines – for simplified calculations with $<s|z^2|s> = <a|z^2|a> = 0$ [2] and the solid ones – for exact calculations.

Thus the negative MR in samples 475/476 is also due to the depopulation of the upper subband. But the upper hole subbands may have a complicated structure with additional lateral extrema at $k_\parallel \neq 0$ [7], which depends on the potential well profile and the uniaxial stress in the layer. This structure should manifest in $\rho(B_\parallel)$ when the Fermi level scans it while moving out from the subband in increasing parallel field. Thus, our results indicate that in samples 475/476 with self-formed DQW profile the structure of the third subband edge is more pronounced than that of the second subband in samples 1123, 1125 with more narrow layers.

**Acknowledgements**

The work is supported by RFBR, projects 02-02-16401, 01-02-17685 and RAS program "1. Low-dimensional quantum nanostructures".

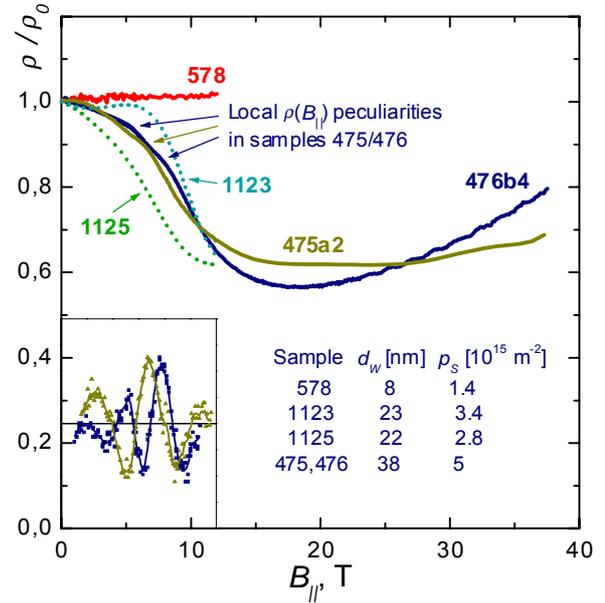

Fig.4. Magnetoresistance of Ge/p-GeSi samples. Insert: MR of samples 475-476 after subtraction of the polynomial fit to the monotonic background.

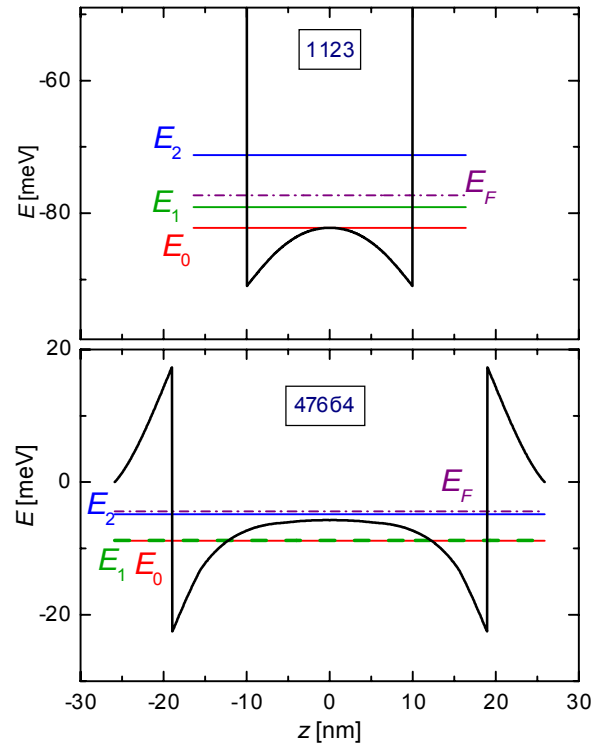

Fig.5. Potential profiles and energy levels for Ge/GeSi QW. In the wide well of sample 476 tow lowest levels merge in the self-formed DQW profile and the third level is slightly populated.